# Energy spectrum of cascade showers generated by cosmic ray muons in water

R.P. Kokoulin, N.S. Barbashina, A.G. Bogdanov, V.D. Burtsev, D.V. Chernov, S.S. Khokhlov, V.A. Khomyakov, V.V. Kindin, K.G. Kompaniets, A.A. Petrukhin, V.V. Shutenko, I.I. Yashin
*National Research Nuclear University MEPhI (Moscow Engineering Physics Institute), Kashirskoe shosse 31, 115409 Moscow, Russia*

The spatial distribution of Cherenkov radiation from cascade showers generated by muons in water has been measured with Cherenkov water calorimeter (CWC) NEVOD. This result allowed to improve the techniques of treating cascade showers with unknown axes by means of CWC response analysis. The techniques of selecting the events with high energy cascade showers and reconstructing their parameters are discussed. Preliminary results of measurements of the spectrum of cascade showers in the energy range 100 GeV – 20 TeV generated by cosmic ray muons at large zenith angles and their comparison with expectation are presented.

## 1. INTRODUCTION

The Cherenkov water calorimeter (CWC) of the Experimental complex NEVOD [1] having a size of $9\times9\times26$ m$^3$ is equipped with a dense spatial lattice of 91 quasi-spherical modules (QSM). QSMs are placed with a step of 2.5 m along the detector, 2 m across it and 2 m in the vertical direction (figure 1).

Each QSM has 6 FEU-200 PMTs with flat photocathodes which are directed along the axes of the orthogonal coordinate system. Such design of the module allows to measure the intensity of Cherenkov radiation from charged relativistic particles with almost equal sensitivity for all directions of incident light; it also allows to determine the light direction. A density of the lattice of QSMs and a wide dynamic range of registered signals (1-10$^5$ photoelectrons for each PMT [2]) allow to measure the spatial distribution of Cherenkov radiation from cascade showers generated in the CWC. The detalization of 0.5 m in this measurement has been achieved (it is a minimum value because of the size of QSM).

These measurements were conducted for events with cascade showers generated by nearly horizontal muons in the sensitive volume of the detector. Muon's track in each event was determined according to data of the coordinate-tracking detector DECOR [3]. Muons crossing the supermodules of DECOR placed in the galleries along two short sides of the CWC were selected. Muons in such events have an average energy of 100 GeV and zenith angles in the range of 85-90°. Track of the muon was assumed to coincide with the axis of the shower, because the sagitta of its trajectory at the length of 26 m is less than 1 cm.

The reconstruction of parameters of cascade showers in these events was made by means of the method [4] using the assumption, that all the cascade particles move close to the axis of the shower and emit photons at the angle of about 42°. To measure the cascade curve of the shower, its axis was divided into bins of one radiation length each (about 36.1 g/cm$^2$ for water). The responses of PMTs that "see" the radiation from the bin at the Cherenkov angle were examined. The response of each such PMT was converted into the number of charged particles corresponding to the bin. In the case of several PMTs, the results were averaged taking into account the errors. Energy of each shower and its point of maximum were reconstructed by fitting the experimental cascade curve with one-dimensional analytical dependence of the number of particles on the depth [5].

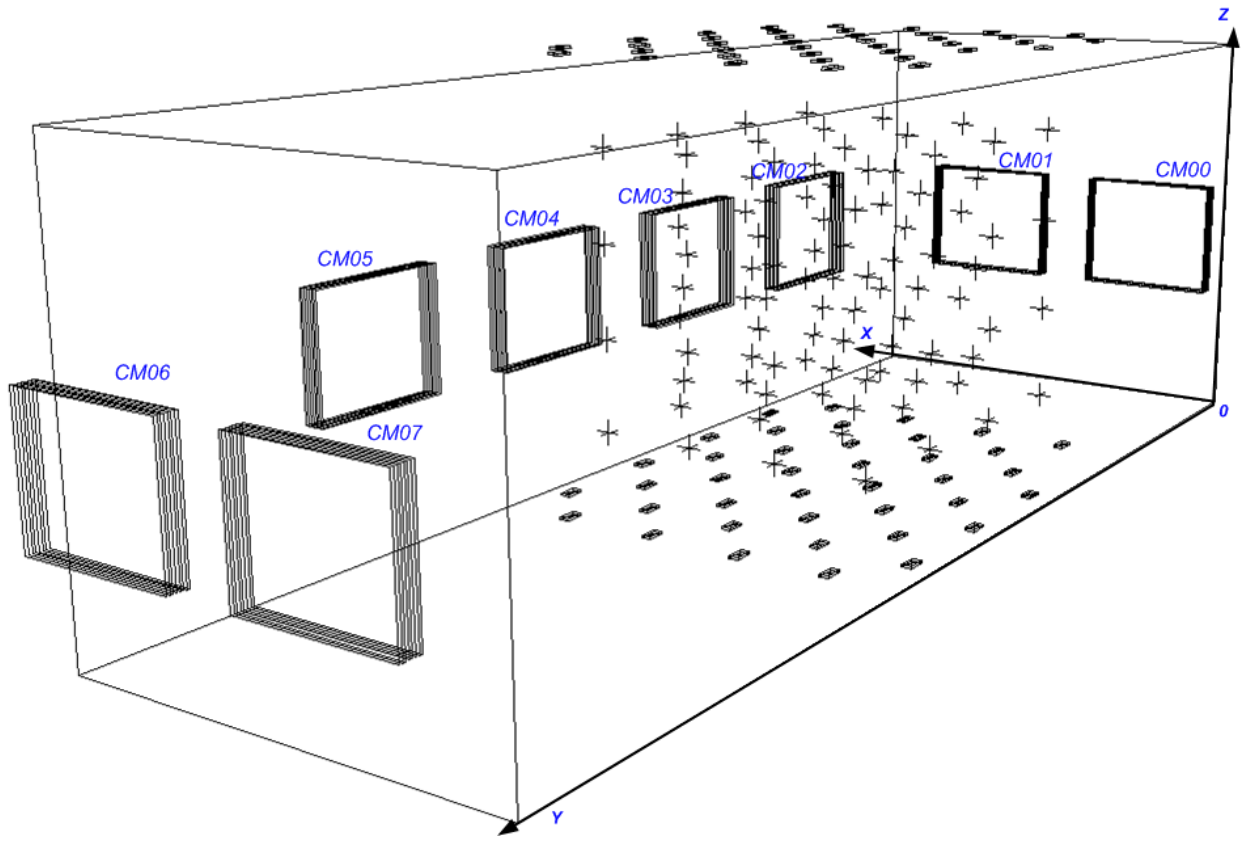

Figure 1: Experimental complex NEVOD-DECOR. Small crosses represent QSMs; big rectangles are DECOR supermodules

## 2. SPATIAL DISTRIBUTION OF THE CHERENKOV RADIATION FROM SHOWERS IN WATER

The spatial distribution of the Cherenkov radiation was measured for cascades with reconstructed energies of 100-500 GeV. Such a narrow energy range was needed to reduce a distortion associated with a longitudinal size of showers (for cascade showers with energies 100 GeV and 500 GeV the distances between the point of generation and the point of maximum of the cascade differ by less than half a meter). 522 cascades were selected among the data of experimental series 2013-2015 with "live time" of about 12 thousand hours. The dependence of QSM response on its distance from the axis and on the depth along the axis was measured. The response of QSM was calculated as the square root of squared amplitudes of all its PMTs (this value almost does not depend on the direction of light [6]), normalized by the energy of the shower:

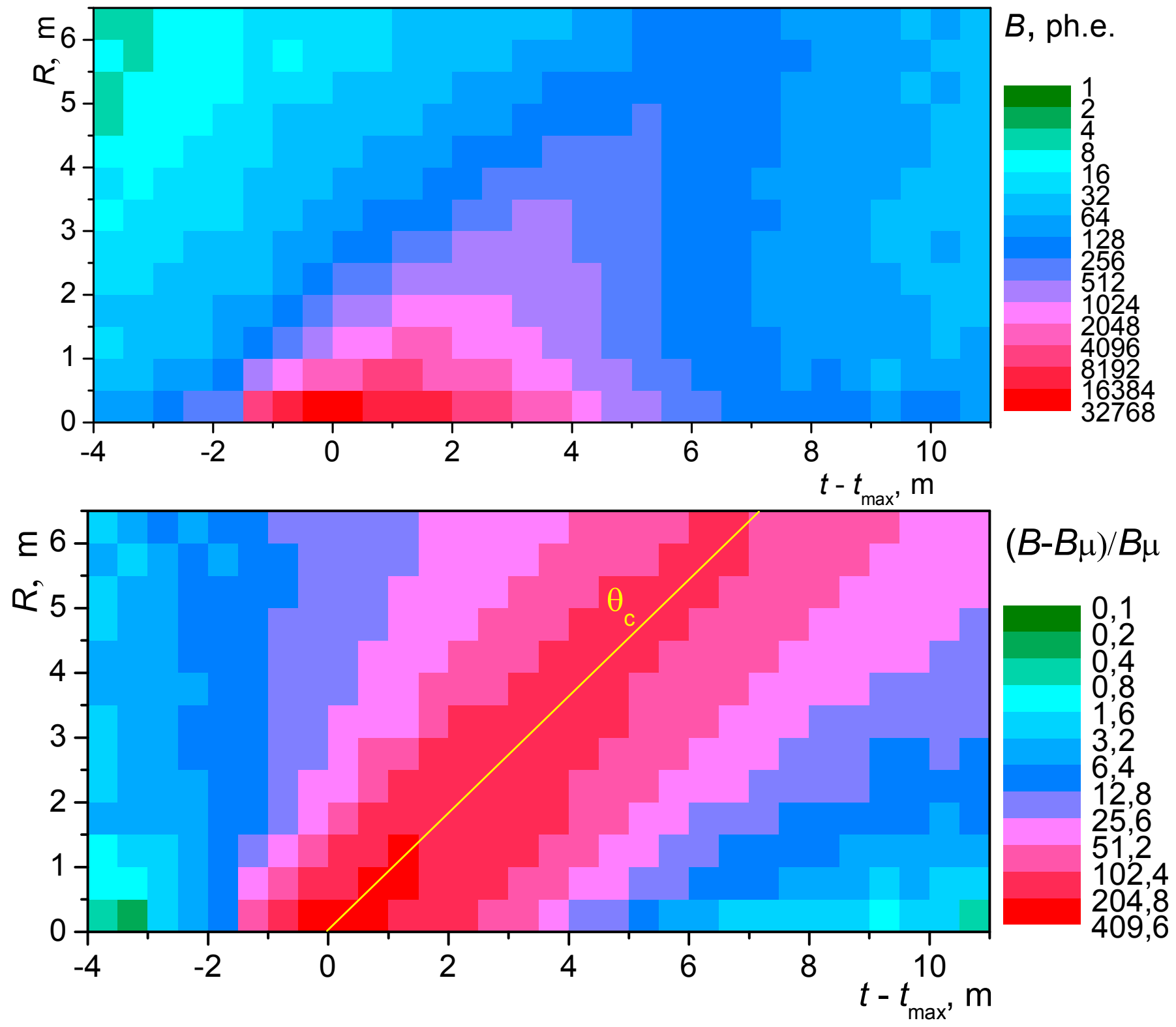

Figure 2. Spatial distribution of Cherenkov radiation from cascades. Bottom frame shows the results with compensation of light attenuation

$$B = \frac{\varepsilon_0}{\varepsilon}\sqrt{\sum A_i^2},$$

where $\varepsilon_0$ = 200 GeV (value close to the average energy of showers in the sample); $\varepsilon$ – reconstructed energy of the shower in the event; $A_i$ – amplitude of the signal of $i$-th PMT in the QSM (in units of photoelectrons, ph.e.).

The concluding dependence of $B$ on the depth and on the distance from the axis was calculated by averaging responses for all events in the sample. The zero point on the axis of the depth corresponds to the point of maximum of each cascade. The obtained dependence is shown in figure 2 (top). It is well seen that the light from the shower spreads in all directions. However, there is a predominant direction of the light cone that corresponds to the front hemisphere defined by the direction of the shower. It is also seen that the region of the highest energy deposit is well-localized. This property was used in the development of the criteria for selection of the cascades among the events with high energy deposit in the CWC.

Attenuation of the Cherenkov radiation and its divergence as a result of multiple scattering of cascade particles are well seen in the obtained dependence. The energy distribution of the cascade particles also affects the character of the dependence, but it is not significant and can be neglected for high energy cascades.

Taking into account that the parameters of light attenuation in water are the same for cascade showers and for single muons, the ratio of light intensity $B$ measured for cascade showers to the intensity $B_\mu$ measured for single muons was considered. It allows to compensate the attenuation of Cherenkov light and to study more efficiently the angular distribution of cascade particles. The spatial distribution of this value is shown in figure 2 (bottom).

The figure shows the change in relative intensity of Cherenkov light depending on the direction defined by the angle with the axis of the shower. It is essential that the boundaries between the areas of different intensity are almost straight lines, which diverge fan-shaped. This confirms the assumption that the factor of light attenuation was practically excluded. The part of the shower near its maximum that accounts for main energy release has a longitudinal size of about two meters.

It is seen that maximums of the longitudinal profiles shift along the axis in accordance with Cherenkov angle. The longitudinal profiles of spatial distribution of $(B\text{-}B_\mu)/B_\mu$ value for different distances from the axis are presented in figure 3 (left). The widening of the longitudinal profiles and decreasing of the maximum values are observed. This is a consequence of the divergence of Cherenkov light because of the scattering of cascade particles. Since the effect of light attenuation is mainly excluded, the areas under the curves of the profiles representing the value of energy of the shower are close to each other. In this case, the profile corresponding to the smallest distance from the shower axis (where the divergence of Cherenkov light is not significant) represents the average cascade curve.

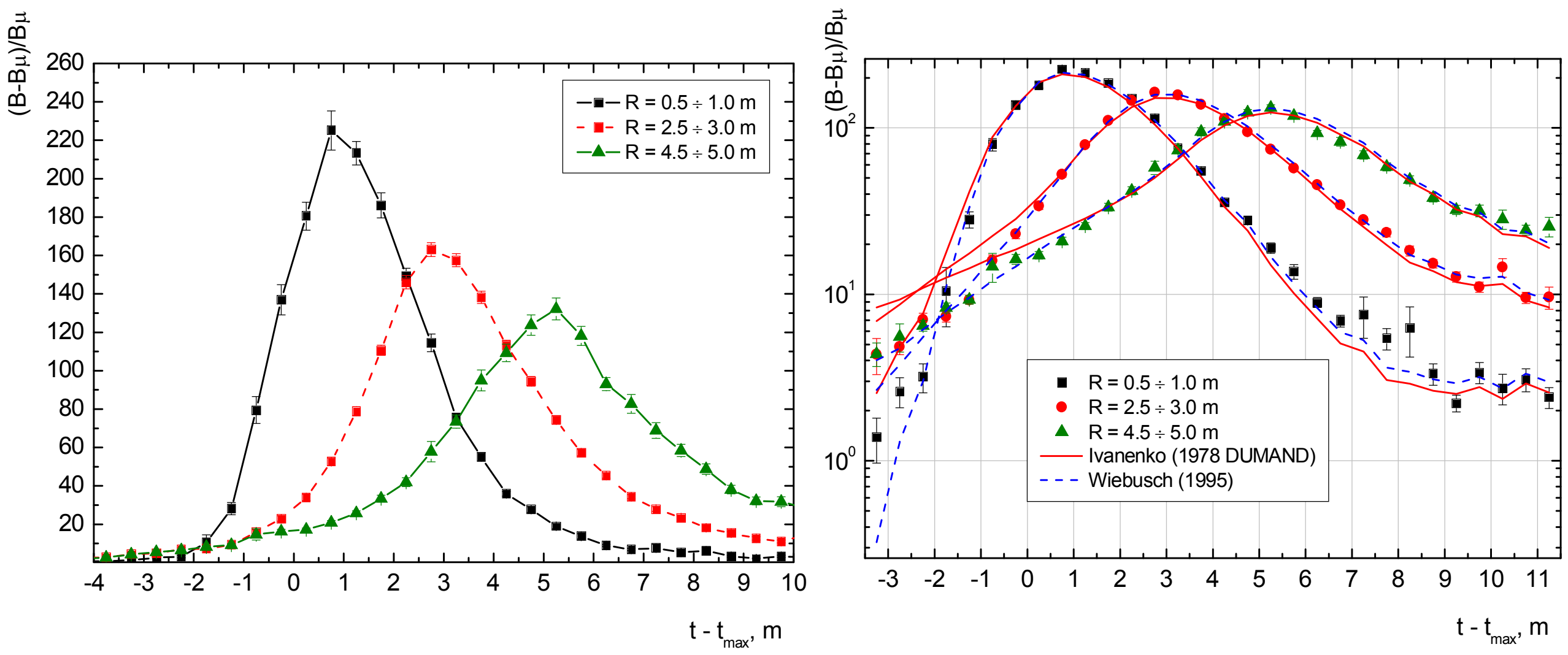

Figure 3. Longitudinal profiles of spatial distribution of the Cherenkov radiation from cascades: experimental data (left), comparison with calculations (right)

## 3. COMPARISON WITH CALCULATIONS

The spatial distributions of the Cherenkov radiation for two models of angular distribution of cascade particles were calculated. First of these models is the analytical calculation on the basis of cascade equations performed by the group of I.P. Ivanenko [7]. The second one is the approximation (constructed by the IceCube collaboration [8]) of the results of modelling in GEANT 3 by C. Wiebusch [9]. The results of comparison of experimental data with calculations are shown in figure 3 (right), where the longitudinal profiles for three distances from the axis are presented.

Picture shows that in the region of maximum of the shower and on the decline of the dependences both models satisfactorily agree with experiment. However on the rising parts of the graphs that correspond to the backward direction of light the model of angular distribution of cascade electrons obtained by C. Wiebusch is preferable.

## 4. ENERGY SPECTRUM OF CASCADE SHOWERS

On the basis of obtained results the algorithm of iterative reconstruction of parameters of cascade showers by means of CWC response was developed. In this algorithm, C. Wiebusch model of angular distribution of cascade particles was chosen for calculation of expected responses of QSMs. Algorithm was tested on the showers with known axes determined using DECOR data. The average error of reconstruction of direction of the axis is about 6°.

The criteria for selecting the cascade showers from the events with high energy deposit (condition of triggering of 60 of 91 QSMs of the CWC) were developed. The main criterion is based on the property that the QSMs with highest responses form a compact cluster in the events with cascade shower (the RMS radius of the cluster < 2.3 m). The cascades with reconstructed zenith angles more than 55° were selected in order to separate showers generated mainly by muons.

Finally, about 200 thousand showers with reconstructed energies above 100 GeV were selected among the data of the experimental series with a duration of about 12 thousand hours. The differential energy spectrum of showers in the range of 0.1 to 20 TeV is shown in figure 4 along with the experimental spectrum for cascades with axes defined using the DECOR detector [10]. Calculations of expected spectra for different values of the slope γ of generation function of parent pions and kaons are also presented.

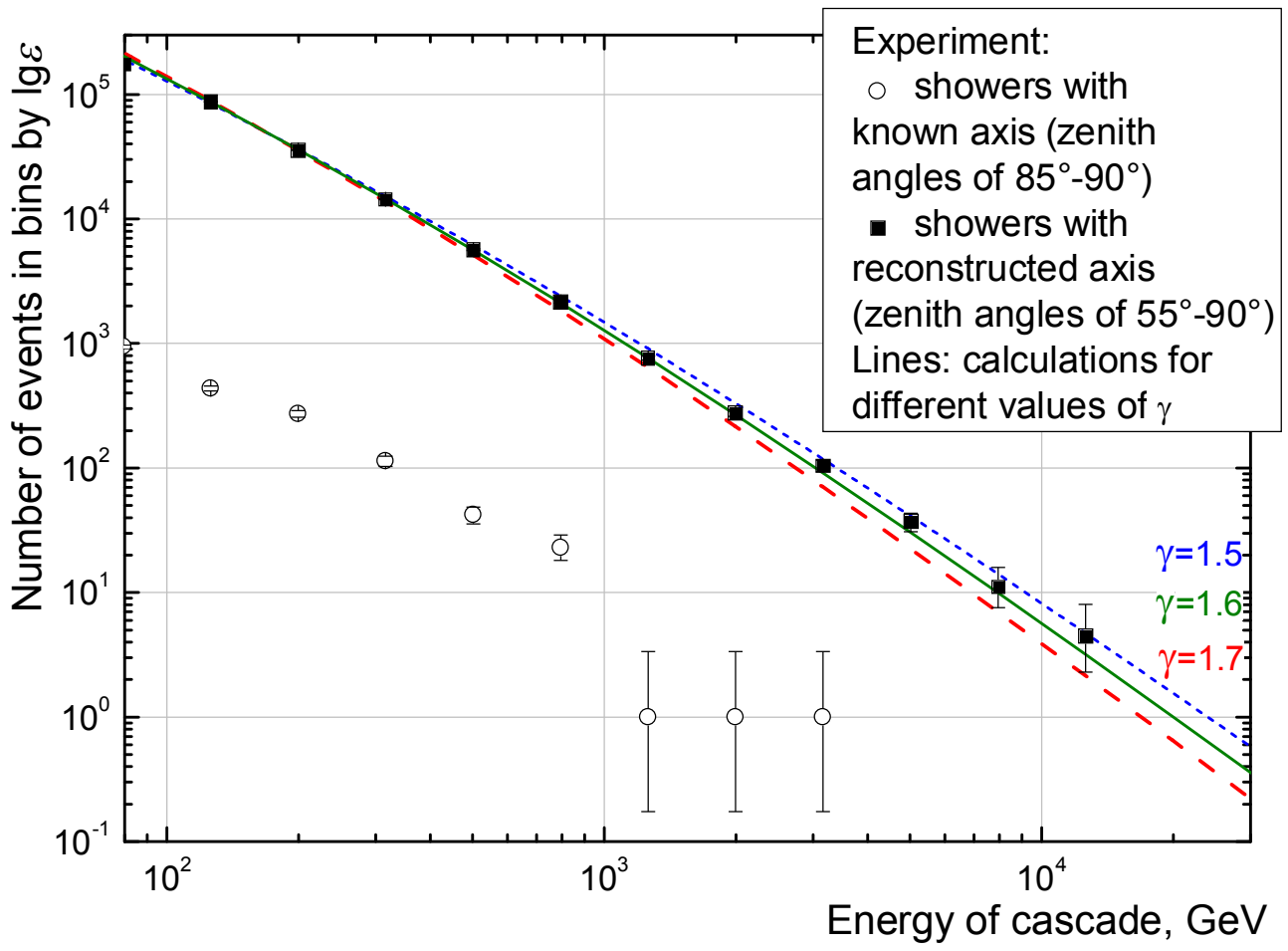

Figure 4. Spectrum of cascade showers initiated by muons in water

The graph shows that the proposed methods for selecting the cascade showers and reconstructing their parameters allowed to increase more than a hundred times the statistics of events with cascade showers; thus the maximum reconstructed energies raised by about one order. This was achieved due to the significant extension of the acceptance.

## CONCLUSIONS

For the first time, the spatial distribution of Cherenkov light from high energy cascade showers generated by muons in water has been experimentally measured. The results demonstrate a good directivity of the light from the cascades at the angle close to the angle of Cherenkov radiation in water (≈ 42°). However, with increasing distance from the axis of the shower the divergence of Cherenkov light due to multiple scattering of cascade particles is clearly visible.

The comparison of experimental data with models of scattering of cascade particles shows that the best consent is observed for the model of angular distribution of cascade particles obtained on the basis of GEANT 3 modeling by C. Wiebusch.

The method of reconstruction of parameters of showers based on the 3D model of spatial distribution of the Cherenkov light is developed. It allowed to increase the statistics of showers, and hence to move to higher energies of showers in the energy spectrum measurements with NEVOD.

## ACKNOWLEDGMENTS

This work was performed at the Unique Scientific Facility “Experimental complex NEVOD”. It was supported by the Russian Ministry of Education and Science (contract RFMEFI59114X0002, MEPhI Academic Excellence Project 02.a03.21.0005 and government task) and Russian Foundation for Basic Research, grant 15-02-07763.